\documentclass{iau} 
\usepackage{graphicx}

\title[Gas accretion in Milky Way-like galaxies]
{How does the stellar disk of the \\ Milky Way get its gas?}

\author[Sebasti\'an E. Nuza et al.]   
{Sebasti\'an E. Nuza$^{1,2}$, Cristina Chiappini$^3$, Cecilia Scannapieco$^2$, Ivan Minchev$^3$, Marie Martig$^4$
 \and Thiago C. Junqueira$^3$}

\affiliation{$^1$Instituto de Astronom\'{\i}a y F\'{\i}sica del Espacio (IAFE, CONICET-UBA), \\ 
CC 67, Suc. 28, 1428 Buenos Aires, Argentina \\ email: {\tt snuza@iafe.uba.ar} \\[\affilskip]
$^2$Facultad de Ciencias Exactas y Naturales (FCEyN), Universidad de Buenos Aires (UBA), \\ 
Buenos Aires, Argentina \\[\affilskip]
$^3$Leibniz-Institut f\"ur Astrophysik Potsdam (AIP), \\ 
An der Sternwarte 16, D-14482 Potsdam, Germany \\[\affilskip]
$^4$Astrophysics Research Institute, Liverpool John Moores University, \\ 
146 Brownlow Hill, Liverpool L3 5RF, UK 
}

\pubyear{2017}
\volume{334}  
\jname{Rediscovering our Galaxy}
\editors{C. Chiappini, I. Minchev, E. Starkenburg \& M. Valentini, eds.}
\begin{document}

\maketitle

\begin{abstract}
In chemodynamical evolution models it is usually assumed that the Milky Way galaxy forms 
from the {\it inside-out} implying that gas inflows onto the 
disk decrease with galactocentric distance. Similarly, to reproduce 
differences between chemical abundances of the thick disk and bulge 
with respect to those of the thin disk, higher accretion fluxes 
at early times are postulated. By using a suite of Milky Way-like galaxies extracted from 
cosmological simulations, we investigate the accretion of gas on the simulated stellar 
disks during their whole evolution. In general, we find that the picture outlined 
above holds, although the detailed behavior depends on the assembly history of the Galaxy 
and the complexities inherent to the physics of galaxy formation. 
\end{abstract}

\firstsection 

\section{Introduction}

A fundamental question in galaxy formation is how galaxies accrete their gas and its 
connection with star formation. Three main channels are believed to be responsible 
of bringing gas to the galaxies, namely: interactions with 
other systems, either by mergers or tidal effects; filamentary accretion 
from the intergalactic medium, and {\it quasi-spherical} accretion of shock-heated material 
in the galactic halo (e.g., \cite[Putman et al. 2012]{Putman_etal12}). 
Provided gas penetrates into the central regions of galactic halos and cools down, 
stellar disks can be feeded with material potentially available for star formation.
Decades ago chemical evolution models (CEMs) have been proposed as a way to study the production of 
chemical elements by subsequent stellar generations which, in turn, can provide insights 
on the required gas accretion and star formation history of galaxies.

The Milky Way is a benchmark for our understanding of galaxy formation 
and evolution due to the wealth of data now available. In general, 
CEMs are quite successful in reproducing the mean chemical patterns of stars 
in the solar neighborhood (\cite[e.g., Chiappini et al. 2001]{Chiappini_etal01}) 
and the time evolution of the thin disk metallicity gradients 
(\cite[Anders et al. 2017]{Anders_etal17}). Several extensions to classical CEMs can be done, 
considering, for instance, inhomogeneous models designed to reproduce the spread in chemical abundances 
(\cite[e.g., Cescutti 2008; Cescutti \& Chiappini 2014]{Cescutti08,Cescutti_etal14}), and chemodynamical 
models to include stellar dynamics information (\cite[Minchev et al. 2013, 2014]{MMC13,MMC14}). 
These models, however, rely on a handful of key assumptions which are usually not very well 
constrained. One of such assumptions is the idea that spirals like our own Galaxy, and in particular 
their thin disk component, are assembled from the {\it inside-out}. Generally speaking, this means that 
gas inflows onto the galaxies are larger in the inner regions. Another key assumption is that the inflows have 
to be significantly stronger at early times, leading, afterwards, to a more gentle gas accretion history 
where stellar disks have enough time to grow as a result of the formation of younger stellar 
populations (\cite[Chiappini et al. 1997]{Chiappini_etal97}). 

In this contributed paper, we discuss 
these ideas by analyzing a set of Milky Way-like galaxies extracted from cosmological simulations 
that we use as a proxy for the accretion of material in our own Galaxy. 
In particular, we quantify the inflow of gas onto the simulated stellar disks as a function 
of cosmic time and radius inspired in the framework adopted by CEMs.

\section{The simulations}

We use two simulated Milky Way-like systems performed within the $\Lambda$CDM cosmology including 
star formation and supernova feedback: the one used in the chemodynamical models of 
\cite[Minchev et al. (2013, 2014)]{MMC13,MMC14}, 
and the Milky Way candidate of \cite[Nuza et al. (2014)]{Nuza_etal14}. 

The former, labelled g106, is selected from the galaxy sample of \cite[Martig et al. (2012)]{Martig_etal12}, 
generated adopting the technique introduced by \cite[Martig et al. (2009)]{Martig_etal09}. 
This technique allows to reduce computational costs at the expense of an approximate 
treatment of cosmological evolution, making possible to achieve higher resolutions. 
Particle masses in this simulation are set to $1.5\times10^4\,$M$_{\odot}$ for gas, $7.5\times10^4\,$M$_{\odot}$ 
for stars and $3.5\times10^5\,$M$_{\odot}$ for dark matter. The target halo was chosen to inhabit a relatively isolated 
environment at $z=0$ with no halo more massive than half of its mass within a region of $2\,$Mpc radius. Therefore, 
by construction, the resulting galaxy candidate displays a quiet merger history. 

The other galaxy, dubbed MW$^{\rm c}$, is extracted from a fully-cosmological simulation of the Local Group (LG) 
carried out within the CLUES project (www.clues-project.org). As a result, MW$^{\rm c}$ inhabits a more realistic 
environment, having today an approaching companion resembling Andromeda at a distance of about 700\,kpc and farther aggregations 
such as Virgo and Coma, among other known large-scale structures in the local Universe. 
This simulation is done with the Smoothed Particle Hydrodynamics 
GADGET-3 code (\cite[Springel et al. 2008]{Springel08}) including additional modules 
for metal-dependent cooling, chemical enrichment, multiphase gas treatment and UV background field 
(see e.g., \cite[Creasey et al. 2015; Scannapieco et al. 2015]{Creasey_etal15,Scannapieco_etal15}). 
In this case, mass resolution is set to $5.5\times10^5\,$M$_{\odot}$ for gas (with two stellar generations) 
and $2.8\times10^6\,$M$_{\odot}$ for dark matter. 

\section{Computing gas fluxes onto the stellar disks}

Following CEMs, we define, at any given time, a cylindrical coordinate system in the plane of the stellar disk. 
The orientation of the latter is obtained by computing the angular momentum vector of star particles in 
each galaxy. Gas particle trajectories are approximated computing the acceleration of matched particles 
between neighboring simulation snapshots. To accommodate the disk, and set the outer boundaries for the flux 
calculations, we use a cylinder of radius $R_{\rm d}=50\,$kpc and scale height $|z_{\rm d}|=2\,$kpc centered 
on the stellar disk midplane. Then, the cylinder is sliced in radial bins and gas 
fluxes accross surface boundaries are computed in the radial and vertical directions. Here, 
we only focus on the vertical infall flux onto the stellar disk leaving to a future study a more thorough analysis 
(Nuza et al., in preparation). 

\begin{figure}
 \hspace*{-0.5cm}
 \includegraphics[width=7.2cm]{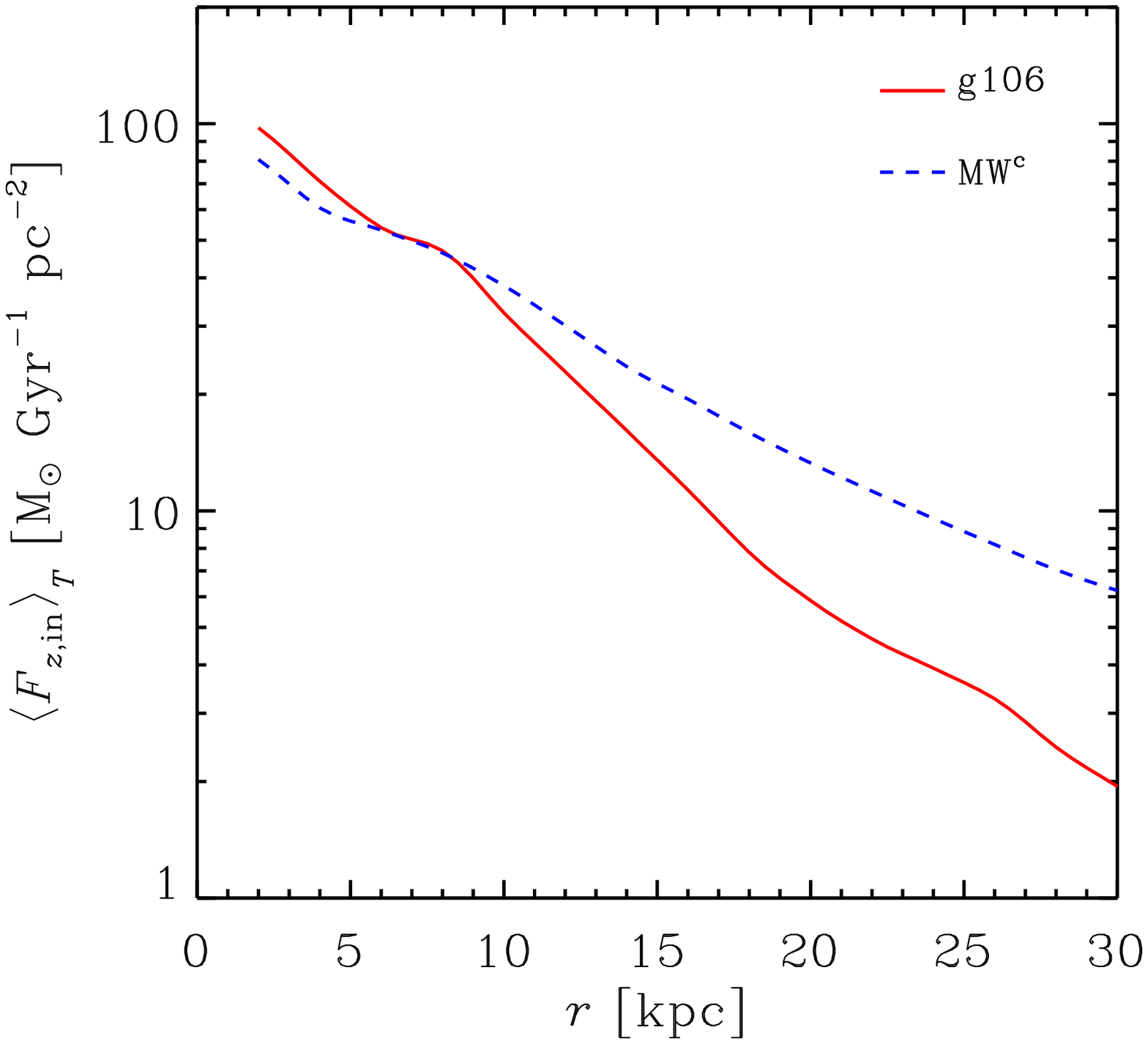}
 \hspace*{-0.5cm}
 \includegraphics[width=7.2cm]{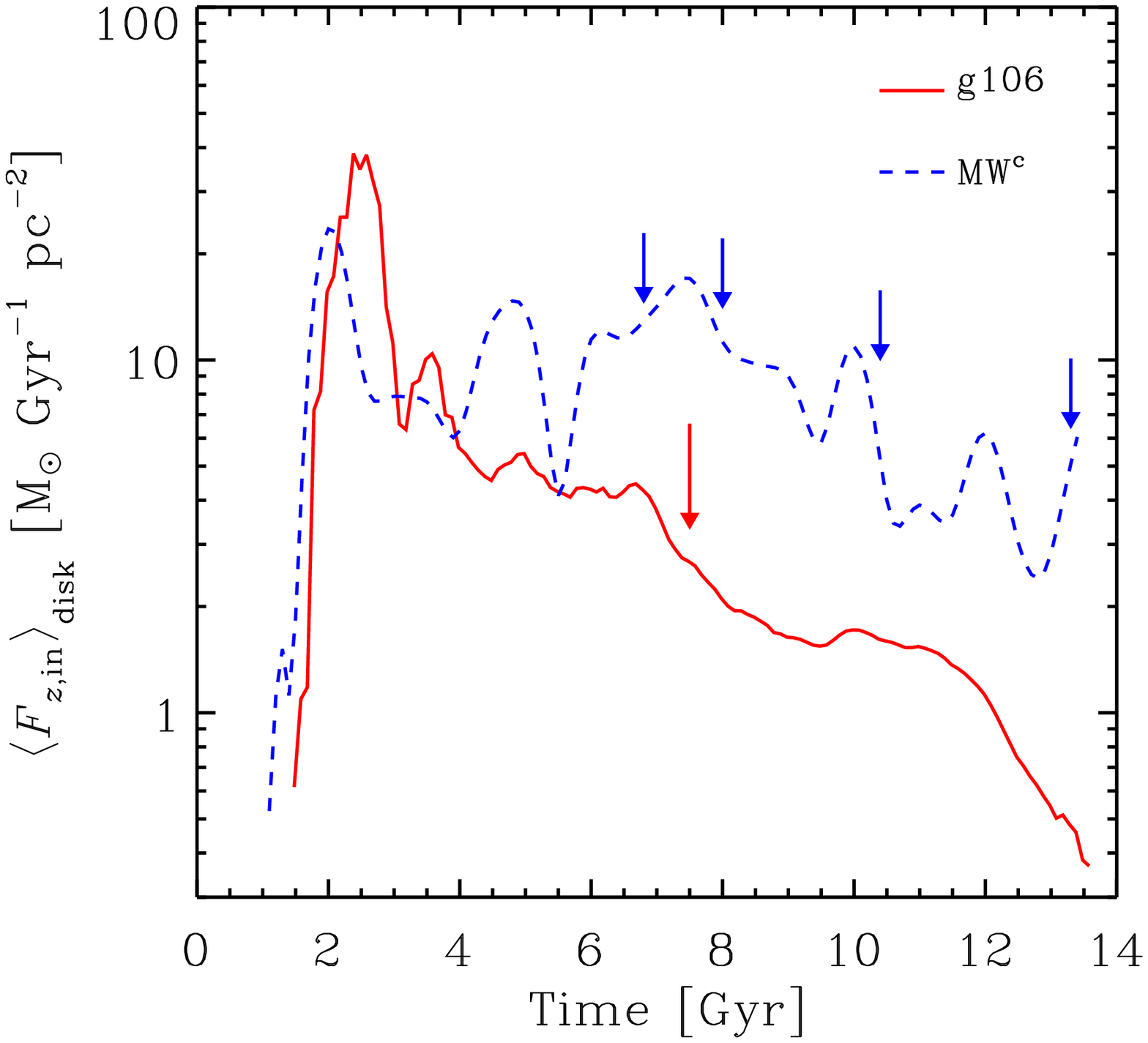} 
  \caption{Mean vertical gas infall onto the simulated Milky Way-like stellar disks. {\it Left:} Mass inflow 
  per unit area {\it averaged} over cosmic time versus radius. {\it Right:} Mass accretion rate {\it averaged} 
  over disk area versus time. Large (small) arrows indicate major (intermediate) mergers 
  as defined in \cite[Scannapieco et al. (2015)]{Scannapieco_etal15}. 
  }
 \label{fig1}
\end{figure}

Figure~\ref{fig1} shows the {\it mean} vertical infall flux of gas onto the disk for the two simulated 
galaxies. The left-hand panel corresponds to the infall mass per unit area for different radii 
{\it averaged} over the entire evolution history, whereas the right-hand panel corresponds to the mass accretion rate 
as a function of time {\it averaged} over the whole disk area. Both simulated galaxies show a behavior consistent with the 
{\it inside-out} formation scenario, where the inner regions of the disk get more of the accreting material. The time-averaged 
infall versus radius displays an exponential downfall towards larger distances. The same behavior 
is still seen at radii well beyond typical stellar disk sizes, suggesting a continuous -- albeit low -- infall of material 
directly feeding the disks. The MW$^{\rm c}$ galaxy shows a flatter decline with radius than g106 due to 
the more active merger history taking place in the LG simulation. Depending on their strength and/or orbital parameters, 
mergers can significantly affect the growth of stellar disks either by completely/partially destroying them, or by bringing 
additional gas leading to star formation.

The radially-integrated infall versus cosmic time shows a downturn towards the present epoch for both galaxies. 
However, g106 displays a much more pronounced decline than MW$^{\rm c}$. As mentioned before, this is a 
direct consequence of the dynamically significant mergers present at later times for MW$^{\rm c}$ (see small arrows in Figure~\ref{fig1}), 
which are revealed by the relative height of the {\it infall peaks} seen at different times. Most of these peaks are related 
to accretion flows induced by interactions/mergers with satellite galaxies. 

Besides the particular assembly history of each model galaxy, we find that, in general, the infall flux of gas onto 
the disk is higher at early times; followed by a more gentle accretion period after the first few Gyr of evolution. 
This scenario is in broad agreement with conclusions of CEMs for the Milky Way which seem to indicate that gas infall 
was more violent in the past than what it is today (e.g., \cite[Chiappini et al. 1997]{Chiappini_etal97}). 
From this plot, however, it is clear that a simple accretion 
law cannot encompass all the details of the complex evolution history of the Galaxy. This is fairly evident 
in the case of MW$^{\rm c}$ but the same can be concluded for g106, despite its more regular decline 
due to the more quiet evolution. 
Nevertheless, when investigating {\it mean} abundance patterns with CEMs, an {\it average} description of the 
infall may be justified. 

\begin{figure}
\begin{center}
 \includegraphics[width=8.7cm]{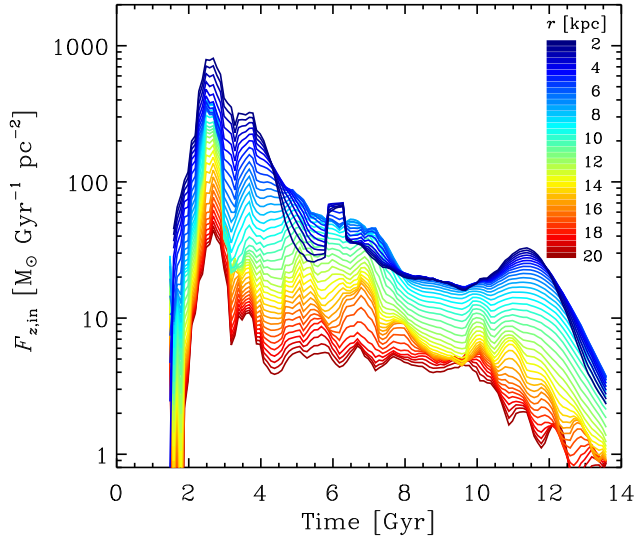} 
\end{center}
  \caption{Vertical infall flux of gas onto the stellar disk of the g106 simulated galaxy versus 
  cosmic time for different distances from the galactic center. The height of each annulus is $4\,$kpc.
  }
   \label{fig2}
\end{figure}

Figure~\ref{fig2} shows the vertical infall flux as a function of cosmic time for our best-resolved Milky 
Way candidate, g106, split in radial bins. In this way, the infall contribution at each annulus in the plane 
of the stellar disk can be cleary seen. As one moves away from the galactic center, the amount of gas accreted 
per unit time and area decreases in a way that is broadly consistent with the already mentioned {\it inside-out} 
formation scenario. Interestingly, for each annulus, the infall flux declines 
{\it exponentially}, in average. This is specially true at late times, where a more quiet 
decline with typical timescales of the order of $5-7\,$Gyr is observed. At early times, 
the downturn is much more pronounced and occurs in timescales of $\sim$$1\,$Gyr. 
These two different accretion episodes might shed light on the formation mechanisms of the thin 
and thick disks (and bulge), as suggested in the two-infall model of 
\cite[Chiappini et al. (1997)]{Chiappini_etal97}, although the discontinuity seen in chemical 
space does suggest a gap in star formation rate (which could be produced by a large merger event, 
in agreement with the results presented by A. Wetzel at this meeting).

\end{document}